\documentstyle[12pt]{article}
\input epsf

\input epsf

\begin{document}

\renewcommand{\thesection}{\arabic{section}.} 
\renewcommand{\theequation}{\thesection \arabic{equation}}
\newcommand{\scs}{\setcounter{equation}{0} \setcounter{section}}
\def\req#1{(\ref{#1})}
\newcommand{\be}{\begin{equation}} \newcommand{\ee}{\end{equation}} 
\newcommand{\ba}{\begin{eqnarray}} \newcommand{\ea}{\end{eqnarray}} 
\newcommand{\la}{\label} \newcommand{\nb}{\normalsize\bf} 
\newcommand{\lb}{\large\bf} \newcommand{\vol}{\hbox{Vol}}
\newcommand{\bb} {\bibitem} \newcommand{\np} {{\it Nucl. Phys. }} 
\newcommand{\pl} {{\it Phys. Lett. }} 
\newcommand{\pr} {{\it Phys. Rev. }} \newcommand{\mpl} {{\it Mod. Phys. Lett. }}
\newcommand{\sg}{{\sqrt g}} \newcommand{\sqhat}{{\sqrt{\hat g}}}
\newcommand{\sqphi}{{\sqrt{\hat g}} e^\phi} 
\newcommand{\sqalpha}{{\sqrt{\hat g}}e^{\alpha\phi}}
\newcommand{\tp}{\cos px\ e^{(p-{\sqrt2})\phi}} \newcommand{\stwo}{{\sqrt2}}
\newcommand{\tr}{\hbox{tr}}

\begin{titlepage}
\renewcommand{\thefootnote}{\fnsymbol{footnote}}

\hfill BUTP-97/20

\hfill hep-th/9707225

\vspace{.4truein}
\begin{center}
 {\LARGE Wrapped M-branes and three-dimensional topologies}
 \end{center}
\vspace{.7truein}

 \begin{center}

 Christof Schmidhuber\footnote{christof@butp.unibe.ch; work supported by Schweizerischer Nationalfonds.}

 \vskip5mm

 {\it Institute of Theoretical Physics, Sidlerstr. 5, 3012 Bern, Switzerland}

 \end{center}

\vspace{.7truein}
\begin{abstract}
The three-dimensional topologies of the membrane of M-theory
can be constructed by performing Dehn surgery along knot lines.
We investigate membranes wrapped around a circle and the correponding subset of 
topologies (Seifert manifolds).
The knot lines are interpreted as magnetic 
flux tubes in an XY model coupled to Maxwell theory. 
In this model the eleventh dimension of M-theory gets ``eaten'' 
by the world-brane metric. 
There is argued to be a 
second-order phase transition at a critical value of the string coupling constant. 
The topology fluctuations that correspond to the knot lines are irrelevant in one phase while they condense in the other phase.
%It is also suggested that the M-brane ``wraps'' because
%momentum in the compact world-brane direction is confined.

\end{abstract}
 \renewcommand{\thefootnote}{\arabic{footnote}}
 \setcounter{footnote}{0}
\end{titlepage}

{}\section*{1. Introduction}\scs{1}

It has become clear recently that
superstring theory contains not only dynamical
strings (1-branes) but also dynamical $p$-branes with $p>1$ (see, e.g., \cite{hulltown}).
 Many of them arise as Dirichlet-branes,
which are stringy generalizations of solitons of field theory \cite{polchinski}.

In particular, type IIA string theory in ten dimensions
contains a Dirichlet two-brane. The world-brane fields that live on it
are the space-time coordinates $x^\mu$ with $\mu\in\{1,...,10\}$ as well as a world-brane U(1) gauge field. Its dual is
a scalar $x_{11}$ which plays the role of an eleventh embedding dimension \cite{dufflu}. 
$x_{11}$  is compact and the
compactification radius (as measured in the eleven-dimensional metric) 
is related to the string coupling constant $\kappa$ \cite{wittendyn}:
\ba x_{11}\ \equiv\ x_{11}+2\pi R\ ,\ \ \ R\sim\kappa^{2\over3}\ .\la{anna}\ea
The world-brane action of the two-brane is the eleven-dimensional supermembrane action
\cite{townsend,schmidhuber} of \cite{bst}. In Nambu-Goto form,
its bosonic part is (setting the three-form to zero):
\ba
S\ \sim\ \int d^3\sigma\ {\sqrt{\det\ G_{ij}}}\ .
\la{bettina}\ea
Here $i,j\in \{1,2,3\}$ and the world-brane signature is taken to be Euclidean.
$G_{ij}$ is the supersymmetric generalization
of the induced world-brane metric $\partial_i\vec x\partial_j\vec x$ 
with $\vec x\equiv(x^\mu,x^{11})$.
We will discuss the bosonic membrane here and the supermembrane in future work.

Most discussions of bosonic and supermembranes have focussed on simple world-brane
 topologies such as $T^3$.
Here we would like to suggest an effect of summing over infinite sets of
three-dimensional topologies: a phase transition at a critical value of $\kappa$.

Three-dimensional topologies are not classified, but there is a procedure, called Dehn surgery, by
which all of them can be constructed (see \cite{rolfsen} for an introduction and \cite{bst} for
some helpful remarks and references). 
One begins with drawing a knot or link (i.e., a set of knots)
on a given manifold. One then cuts out a
tubular neighborhood around each of the knotted lines and glues it back in differently (Fig. 1): 
call $T'$ the torus that is the boundary of the tubular neighborhood, and $T$ the torus that
is the boundary of its complement. Then $T'$ is identified with $T$
by an SL(2,Z) transformation that maps the meridian $M'$ of $T'$ onto $nM+mL$, where $M$ and $L$
are a meridian and a longitude on $T$ and $m,n$ are relatively prime.
All three-dimensional topologies can be constructed by starting with an $S^3$,
drawing all possible links on it, and performing all possible surgeries on them.
(Thus the classification of three-dimensional topologies requires
in particular a classification of links.)

 \begin{figure}[htb]
 \vspace{9pt}
 \vskip7cm
 \epsffile[-60 1 0 0]{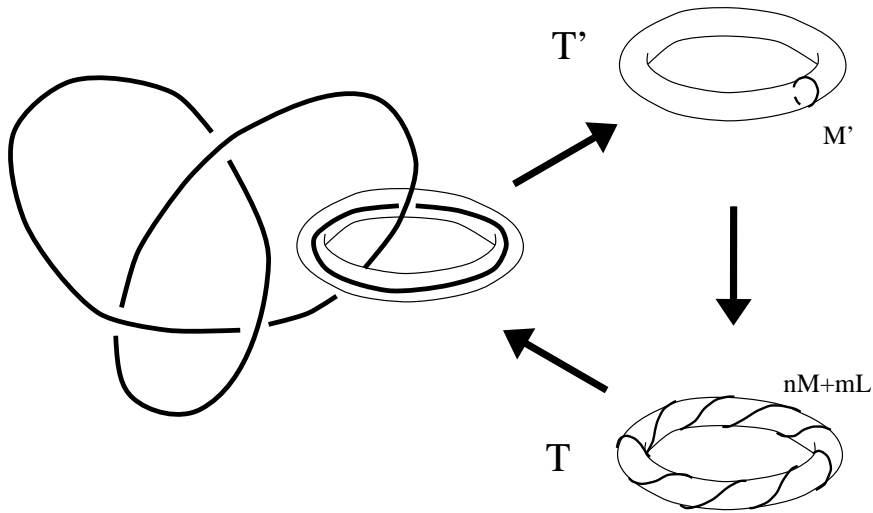}
 \caption{\small Surgery on one of the two knots of a link ($m=2,n=9$)}
 \end{figure}

%We do not attempt to sum over {\it all} topologies 
In this paper we consider
membranes that are ``wrapped'' around the circular $x_{11}$ direction. As will be discussed,
this restricts the possible three-dimensional topologies to the (still very rich)
subclass of ``Seifert manifolds'' \cite{seifert}
 - manifolds that can be foliated by circles. These
can be constructed by starting with $\Sigma\times S^1$, where $\Sigma$ is a (possibly unorientable)
 Riemann surface
of arbirary genus, choosing a ``link'' which consists of an arbitrary number of lines running around
the $S^1$, and performing independent surgery on each line.

As argued in section 2, the description of dynamical wrapped membranes
involves an XY model coupled to Maxwell theory. The Maxwell field originates from
an auxiliary world-brane 
metric and ``eats'' the scalar field $x_{11}$ of the XY model. This model can be seen to contain
magnetic flux tubes.
It is argued in section 3 that these flux tubes precisely correspond to
knot lines on which Dehn surgery is performed. 
From this it is concluded section 4 that
there is a second-order phase transition at a critical string coupling constant $\kappa_c$: for
$\kappa>\kappa_c$, Dehn surgery lines are irrelevant while for
$\kappa<\kappa_c$ they condense.
%The two phases might correspond to type IIA and IIB string theory.

In section 5, it is suggested that the ``wrapping'' of branes can be understood
as confinement of momentum in compact world-brane directions (motivated by a comment in \cite{wittencom}).

{}\section*{2. Wrapped dynamical M-branes}\scs{2}

Let us begin with some general remarks. The world-brane action (\ref{bettina}) of the two-brane is in Nambu-Goto form.
From our experience with one-branes it seems clear that studying only Nambu-Goto actions is clumsy and misses
many important aspects of string theory.
In the one-brane case, the fruitful thing to do is to rewrite the
Nambu-Goto action with the help of an auxiliary $2d$ metric $h_{\alpha\beta}$ as \cite{tucker}
$$\int d^2\sigma{\sqrt{|\partial_\alpha\vec x\partial_\beta\vec x|}}\ \ \ \rightarrow\ \ \ 
\int d^2\sigma\ {\sqrt{h}}h^{\alpha\beta}\partial_\alpha\vec x\partial_\beta\vec x \ .$$
Classically, the saddle point value of the integral over $h_{\alpha\beta}$ reproduces the Nambu-Goto action.
Quantum mechanically, it is important to include in the action
the terms induced in the process of renormalizing this
world-sheet theory of gravity coupled to matter fields $\vec x$ (the ``Polyakov string''). In particular, it is crucial to
include the conformal anomaly term ${\cal R}{1\over \Box}{\cal R}$, where ${\cal R}$ is the two-curvature,
and compute its coefficient
in order to deduce the critical dimension \cite{polyakov}
 or the equations of motion of perturbative string theory \cite{fratsey}.

In trying to mimick this procedure for two-branes one of course meets at least two outstanding problems.
First, (\ref{bettina}) can still be rewritten with the help of a $3d$ metric $h_{ij}$ as
\ba
S\ \sim\ \int d^3\sigma\ {\sqrt{h}}\{h^{ij}\partial_i\vec x\partial_j\vec x \ -\ 1 \}\ .
\la{diana}\ea
But in three dimensions, a counterterm that should be included in the action
-- at least for the bosonic membrane\footnote{It is not clear to the author whether
such a term can be avoided in the case of the supermembrane} -- is the Hilbert-Einstein term
\ba\int {\sqrt{g}}{\cal R}^{(3)}\ .\la{beate}\ea
This leads to 
a nonrenormalizable theory of three-dimensional gravity coupled to matter that does not seem to make 
sense (although three-dimensional gravity without matter is a topological theory \cite{witten3d}).
A second obvious problem is that the sum over three-dimensional topologies seems rather intractible.

The purpose of this paper is to discuss is a truncation of the problem that leads to a
sensible and nice extension of the ``Polyakov string'',\footnote{Like 
the standard string, this extension of it can perhaps be thought of as an
effective description of (some limit of) M(atrix) theory \cite{susskind},
where the M-brane is a bound state of Dirichlet zero-branes. But our conclusions do
not depend on what the M-brane is thought to be composed of.}
and moreover turns out to
lead to a tractible sum over a subset of three-dimensional topologies: Seifert 
manifolds.

The truncation consists of summing only over ``wrapped membrane embeddings'' $\vec x(\sigma)$.
By this the following is meant. We consider closed bosonic Euclidean
membranes $M$ whose (internal) topology is such that they are locally of the form $D\times S^1$, where $D$ is a disc. 
Thus, $M$ must admit a foliation by circles.
The parameter along the $S^1$ is $\sigma_3\in[0,2\pi[$, while the disc 
coordinates are $\sigma_1,\sigma_2$.
In the simplest case,
$M$ might be $\Sigma\times S^1$ where $\Sigma$ is a Riemann surface covered by a set of discs;
but we will be more interested in the general case where $\sigma_3$ cannot be defined globally
(without coordinate singularities).
We now restrict the path integral 
over the target space coordinates $\vec x(\sigma)\equiv(x^\mu(\sigma),x_{11}(\sigma))$ 
in each patch $D\times S^1$ of $M$
to the following configurations.
First, $x_{11}$ changes by $nR$ around the $S^1$ where $n$ is an integer. Second, 
$\vec x$ is otherwise independent of $\sigma_3$:
\ba
x^\mu\ =\ x^\mu(\sigma_1,\sigma_2)\ \ ,\ \ \ x_{11}\ =\ nR\sigma_3\ +\ \tilde x_{11}(\sigma_1,\sigma_2)\ .
\la{clara}\ea
Locally, $\tilde x_{11}$ can be set to zero by a shift of $\sigma_3$. But it will be crucial
that this cannot be done globally for an M-brane of general world-brane topology.

Of course the restriction (\ref{clara}) 
breaks three-dimensional general covariance.
That it is imposed ``by hand'' is unsatisfactory, and in the last section it will be suggested
when and how it might arise dynamically,
namely as confinement of momentum along the $S^1$.
But for now let us simply include this restriction in the definition of the problem and just note that it 
is also tacitly assumed in previous attempts of regarding dynamical strings as wrapped dynamical membranes,
such as \cite{schwarz}.

Actually, the fields $x^\mu(\sigma_1,\sigma_2)$ will mostly be spectators in this paper and
could in principle be dropped. 
If they are dropped, what remains are membranes embedded in one compact dimension $x_{11}$.
This may seem to be a trivial problem if one just looks at the Nambu-Goto action (\ref{bettina})
which is zero in this case.
But as will become apparent, when treated properly the problem turns out to be very rich - 
just like the problem of strings embedded in one or less dimensions is very nontrivial.

Why does restriction (\ref{clara}) make the  
sum over membrane configurations tractible? With this restriction
the induced metric $G_{ij}$ depends only on $\sigma_\alpha$, where $\alpha\in\{1,2\}$, and not on $\sigma_3$:
\ba G_{\alpha\beta} &=& \partial_\alpha\vec x\ \partial_\beta\vec x(\sigma_1,\sigma_2)\\
G_{\alpha3} &=& nR\ \partial_\alpha \tilde x_{11}(\sigma_1,\sigma_2)\\
G_{33} &=& (nR)^2\ .\ea
The action (\ref{bettina}) then becomes a two-dimensional integral over $\Sigma$.
Therefore the auxiliary metric $h_{ij}$ in (\ref{diana}) can be chosen to also depend only on $\sigma_1,\sigma_2$.
This is sufficient for the saddle point of the path integral over $h_{ij}$ at 
\ba h_{ij}(\sigma_1,\sigma_2)\ =\ G_{ij}(\sigma_1,\sigma_2)\equiv\partial_i\vec x\partial_j\vec x\la{ines}\ea
to reproduce the Nambu-Goto action (\ref{bettina}).

Actually, since $G_{33}$ is a constant that does not depend on $x^\mu,\tilde x_{11}$ at all, 
it is not even necessary 
to introduce $h_{33}$ as an independent field; it can in principle
 be set equal to $G_{33}$ from the beginning:
\ba h_{33}\ =\ (nR)^2\ .\la{sima}\ea
In other words, we take the viewpoint that
there are two possibilities to define the theory to be studied:
either we introduce only auxiliary world-brane
fields $h_{\alpha\beta}(\sigma),h_{3\alpha}(\sigma)$ and impose (\ref{sima}).
Or we introduce $L^2(\sigma)=h_{33}(\sigma)$ as an independent field.
Both definitions reduce to the Nambu-Goto membrane
(\ref{bettina}) if the term (\ref{beate}) is ignored.\footnote{This can be compared with
the fact that the Polyakov string reduces to the Nambu-Goto string if the conformal anomaly
term ${\cal R}{1\over\Box}{\cal R}$ is ignored.}
For simplicity we will impose (\ref{sima}) here, leaving the general case for future investigation.

The three-dimensional metric $h_{ij}$ can now
be thought of as being composed of a two-dimensional metric $h_{\alpha\beta}(\sigma)$,
a two-dimensional abelian gauge field $A_\alpha(\sigma)$, and a scalar $L$ which has been set equal to $(nR)$.
The metric and the gauge field are defined, as usual in Kaluza-Klein reduction, by the line element
\ba
(ds)^2\ =\ h^{(2)}_{\alpha\beta}\ d\sigma^\alpha d\sigma^\beta \ +\ (nR)^2(d\sigma_3+A_\alpha d\sigma^\alpha)^2\ ,
\la{elise}\ea
such that $h_{3\alpha}=(nR)^2A_\alpha$.
The kinetic term in (\ref{diana}) for the space-time coordinates $\vec x$ becomes
\ba
  h^{ij}\partial_i\vec x\partial_j \vec x\  \ \ \rightarrow\ \    
  {1\over (nR)^2}(\partial_{3}\vec x)^2\ +\ h^{\alpha\beta}(\partial_\alpha \vec x -A_\alpha\partial_3\vec x )
  (\partial_\beta \vec x -A_\beta\partial_3\vec x )\  .\ea
With ansatz (\ref{clara}), 
   \ba 
  h^{ij}\partial_i\vec x\partial_j \vec x\ -1\ \ =\ \    
  h^{\alpha\beta}\{\partial_\alpha x^\mu 
  \partial_\beta x_\mu \ +\ (\partial_\alpha x_{11} -nRA_\alpha )
  (\partial_\beta x_{11} -nRA_\beta)\}\ .
\ea 
The U(1) gauge transformations for which $A_\alpha$ is the gauge field are translations in $\sigma_3$--direction, and
the associated charge is momentum in $\sigma_3$-direction.
Thus the $x^\mu$ become neutral two-dimensional fields, while $x_{11}$, or rather $e^{ix_{11}/R}$ has charge $n$.

The three-dimensional Hilbert-Einstein action (\ref{beate})
becomes the Maxwell action plus the Euler characteristic of $\Sigma$ (the integral over
the two-curvature ${\cal R}^{(2)}$),
\ba
\int d^3\sigma{\sqrt h} {\cal R}^{(3)} \ \rightarrow\ -2\int d^2\sigma\ {\sqrt{h^{(2)}}}(nR)\ 
\{{1\over4}(nR)^2F_{\alpha\beta}^2
-{\cal R}^{(2)} \}\ ,
\la{erika}\ea
where $F_{\alpha\beta}$ is the field strength of $A_\alpha$.
We will include this term in the action, introducing an undetermined coupling constant $\gamma$
whose value will not be relevant for the conclusions.
Note however that its value is related to the string coupling constant if we regard both
the Maxwell term and ${\cal R}^{(2)}$ as remnants of the three-dimensional Hilbert-Einstein term.
The complete action (\ref{diana}+\ref{erika}) is now\footnote{we assume cancellation of the
conformal anomaly term ${\cal R}{1\over\Box}{\cal R}$ and the absence of a chiral anomaly term $F{1\over\Box}F$
which occurs only in the presence of charged chiral fermions}
\ba
 S&\sim&\int d^2\sigma\sqrt{ h ^{(2)}}\ (nR)\ \{ h^{\alpha\beta}\partial_\alpha x^\mu 
  \partial_\beta x_\mu \ -\ \gamma {\cal R}^{(2)}  \}\la{flavia}\\
&+&\int d^2\sigma\sqrt{ h ^{(2)}}\ (nR)\ \{
 h^{\alpha\beta}(\partial_\alpha x_{11} -nRA_\alpha )
  (\partial_\beta x_{11} -nRA_\beta) \ +\ {\gamma\over4}(nR)^2F_{\alpha\beta}^2 \}\ .
\la{gabi}\ea
The first line represents a standard string with tension $(nR)$ 
(plus the Euler characteristic). 
The main new input from M-theory is the second line. It
is the action of an XY model coupled to Maxwell theory 
in a curved background defined by $h_{\alpha\beta}$.
$x_{11}$ plays the role of the scalar field of the XY model.

This system is coupled to dynamical $2d$ gravity.
The strategy will be to first study the model in flat background $h_{\alpha\beta}=\delta_{\alpha\beta}$.
Afterwards we will discuss the coupling to gravity.
To study the model in flat space, we replace $x_{11}$ by an angle $\Theta$
and redefine $A_\alpha$:
\ba
\Theta\ \equiv\ {x_{11}\over R}\ \ ,
\ \ \ \ \tilde A_\alpha\ \equiv\ nA_\alpha\ .\la{pamela}\ea
Then the (conveniently normalized) action (\ref{gabi}) becomes
\ba
S\ =\ {1\over4\pi}\int d^2\sigma\ \{\ {1\over T}(\partial_\alpha\Theta-\tilde A_\alpha)^2\ +\ {1\over 4e^2}\tilde
F^2\ \}
 \la{kate}
\ea
with ``temperature''
\ba T\ =\ {1\over nR^3}\ \ea
and ``electric charge'' $e$ given by
$$ \ e^2\ =\ {1\over\gamma nR^3}\ .$$
This model might seem to be rather trivial since $\Theta$, i.e. $x_{11}$, can be gauged away locally:
one could say that the eleventh dimension of M-theory is
a Goldstone boson that gets eaten by (part of) the world-brane metric, and what remains is
a standard string in ten dimensions.
However, as discussed in section 4 there exist 
 magnetic flux tubes  that involve vortex lines of $x_{11}$
and will be argued to have interesting effects.

{}\section*{3. Magnetic flux and Seifert manifolds}\scs{3}

Before studying the system (\ref{kate}), let us discuss magnetic flux in the model. 
To this end, we first consider the case $n=1$, i.e. we assume that the membrane wraps once around the
$x_{11}$-direction. Then $\tilde A_\alpha=A_\alpha$.
The total magnetic flux through $\Sigma$ is quantized according to the Dirac rule:
\ba
\int_\Sigma F\ =\ 2\pi\ {m}\ ,\ \ \ m\in Z\ .\la{carola}
\ea
For simplicity, let us first assume that $\Sigma$ is a two-sphere.
The wrapped M-brane can be regarded as the total space of a $U(1)$ fibre bundle
over $\Sigma$, the fibre representing the circle parametrized by $\sigma_3$.
Let us first demonstrate that in the case at hand
this total space is the so-called ``lens space'' $L(m,1)$
(see also, e.g., \cite{hull}.)

To this end we parametrize the two-sphere $\Sigma$ by spherical coordinates
$\theta\in [0,\pi]$ and $\phi\in[0,2\pi]$ with the North pole at $\theta=0$ and 
the South pole at $\theta=\pi$.
If the gauge field is chosen to only have a tangential component, $A_\alpha=A\ (e_\phi)_\alpha$,
with $A\rightarrow0$ at the South pole, 
then there is a Dirac string at the North pole:
\ba A\rightarrow-{m\over \theta}\ \ \ \hbox{as}\ \ \ \theta\rightarrow0\ .\la{julia}\ea
That is, the metric $h_{3\alpha}=(nR)^2A_\alpha$ in (\ref{elise}) is singular at the North pole. 
This singularity in the metric
can be removed by changing coordinates from $(\theta,\phi,\sigma_3)$ to $(\theta',\phi',\sigma_3')$ with
\ba
\theta'&=&\theta\\
\phi'&=&\phi\\
\sigma_3'&=&\sigma_3-m\phi\ .\la{iris}\ea
To leave the line element (\ref{elise}),
\ba
ds^2\ =\ d\theta^2\ +\ \rho^2(\theta)\ d\phi^2\ +\ L^2(d\sigma_3\ +\ A(\theta)\rho(\theta)\ d\phi)^2\ \la{katharina}
\ea
with
$$\rho(\theta)\ =\ \sin\theta\ \ ,\ \ \ L\ =\ nR \ ,$$
invariant, the new gauge field $A'$ must be related to the old one by a large gauge transformation:
$$ A'\ =\ A\ +\ {m\over \rho}\ .$$
$A'$ is then zero at the North pole.
Consider now the little circle around the North pole defined in the new coordinate system by
$${\gamma'}\ \ :\ \ \theta'=\epsilon\ \ ,\ \ \ \sigma_3'=0\ \ ,\ \ \ \phi'\in[0,2\pi[\ .$$
In the old coordinate system it is represented by the closed spiral
$${\gamma}\ \ :\ \ \theta=\epsilon\ \ ,\ \ \ \sigma_3=m\phi\ \ ,\ \ \ \phi\in[0,2\pi[\ .$$

So a three-dimensional topology on which a metric lives whose component $h_{3\alpha}=L^2A_\alpha$ obeys (\ref{carola})
can be constructed as follows:
One takes two solid tori $S=D\times S^1$ and $S'=D'\times S^1$, where $D$ and $D'$ are disks.
$D'$ here represents, e.g., a small disk $\theta\le\epsilon$ 
that is cut out of $\Sigma$ around the North pole, 
and $D$ represents the remainder of $\Sigma$ (with $\theta\ge\epsilon$).
One now identifies the toroidal suface $T$ of $S$ with the toroidal surface $T'$ of $S'$
by a homeomorphism that maps
the meridian ${\gamma'}$ of $T'$ onto the closed spiral line ${\gamma}$ on $T$ (Fig. 2).

This operation of cutting out a tubular neighborhood
of a three-manifold (here of $\Sigma\times S^1$) and gluing it back with a different identification
is a special type of Dehn surgery. 
The topologies constructed in the case described are by definition
the lens spaces $L(m,1)$ (see \cite{rolfsen}).
They differ from $S^2\times S^1$ e.g. in their first homology group, which is $Z_m$.

\begin{figure}[htb]
\vspace{9pt}
\vskip5cm
\epsffile[-70 1 0 0]{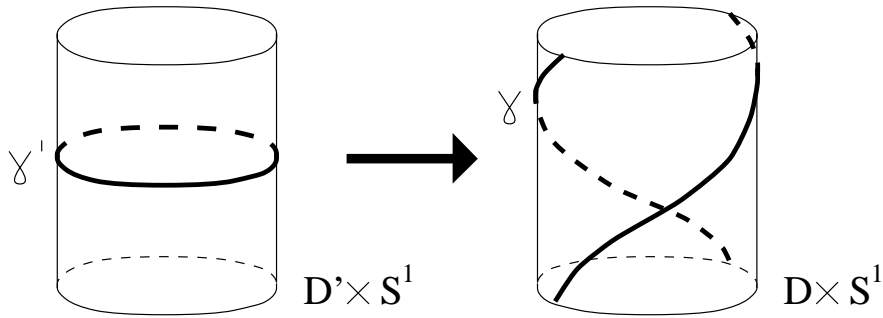}
\caption{\small Surgery with $m=2,n=1$; the bottom and top of the cylinders are identified.}
\end{figure}

Another way of seeing that the magnetic flux (\ref{carola}) changes the topology of the three-manifold
is to consider the curve $\beta$ defined in the old coordinate system by 
$$\beta:\ \ \theta=\epsilon\ ,\ \ \sigma_3=0\ ,\ \ \phi\in[0,2\pi[\ .$$ 
Although in the old coordinate system $\beta$ looks like a little circle
around the North pole, from (\ref{julia}) and metric (\ref{katharina}) one finds that
it retains finite length $2\pi{m}L$ as $\epsilon\rightarrow0$.
Thus it cannot be contracted to a point in the neighborhood of the North pole.
There is however a closed line that {\it is} contractible there. This is the spiral
$\gamma$. Indeed, with metric (\ref{elise}) this spiral
has length $2\pi\epsilon$ and shrinks to a point as $\epsilon\rightarrow0$.

Let us next consider the case of wrapping number $n>1$ in (\ref{clara}). 
In the next section we will encounter field configurations with
fractional magnetic charge 
\ba
\int_\Sigma F\ =\ 2\pi\ {m\over n}\ ,\  \ \ \hbox{with}\ \ \  m,n\in N\ \hbox{and relatively prime}\ .
\la{hanna}\ea
What is their geometric interpretation? Let us pick a
coordinate system in which $A_\alpha$ is regular at the South pole.
Near the North pole, the gauge field then diverges as $A\rightarrow-{m\over n\theta}$.
Analogously to the case $n=1$, it is easy to see from (\ref{katharina})
that there is a closed curve that is contractible near the North pole: the spiral line defined by
$$ {\gamma}\ \ :\ \ \theta=\epsilon\ \ ,\ \ \ n\sigma_3-m\phi=0\ \ ,\ \ \ \phi\in[0,2\pi n[ $$
has length $2\pi n\epsilon$ and shrinks to a point as $\epsilon\rightarrow0$.

What is a local coordinate system $(\theta',\phi',\sigma_3')$ 
in the neighborhood of the North pole in which $A_\alpha$ (and therefore the $3d$ metric) is non-singular?
It is of course not possible to choose $\theta'=\theta,\phi'=\phi$ and $\sigma_3'=\sigma_3-{m\over n}\phi$,
because then $\gamma$ would be mapped onto an $n$-fold cover of the circle $\sigma_3'=0,\theta'=\epsilon,\phi'\in[0,2\pi[.$
Equivalently, this circle would not correspond to a closed line in the old coordinate system.
Instead, one can perform the SL(2,Z) transformation
\ba
\theta'&=&\theta\la{max}\\
\phi'&=&a\phi+ b\sigma_3\\ 
\sigma_3'&=&-m\phi+n\sigma_3\la{john} 
,\ea
where $a,b$ are integers such that $an+bm=1$. The latter condition ensures that the determinant of the map is one,
while the condition $a,b\in Z$ ensures that closed lines are mapped onto closed lines.
It can be seen that different choices of $a,b$ that satisfy both conditions lead to equivalent topologies.

To leave the line element (\ref{katharina}) invariant, the ``large diffeomorphism'' (\ref{max}-\ref{john})
must be accompanied by a transformation
$$\tau\ \rightarrow\ {n\tau+m\over -b\tau+a}\ $$
of the modular parameter
$$\tau\ =\ \rho A\ +\ i{\rho\over L}$$
of the torus that is - for given $\theta$ - parametrized by $z$ and $\phi$.
In the new coordinate system, $A$ is of order $\theta$ and goes to zero at the North pole.
Note that $\rho$ are $L$ also changed in the new coordinate system: $\rho\rightarrow n\rho$ and $L\rightarrow{L\over n}=R$, and therefore ${\rho\over L}\rightarrow n^2{\rho\over L}$ at the North pole. 

The spiral ${\gamma}$ of the old coordinate system
is now indeed mapped one-to-one onto the circle $\gamma'$ in the new coordinate system defined by
$${\cal \gamma'}\ \ :\ \ \theta'=\epsilon\ ,\ \ \ \sigma_3'=0\ \ ,\ \ \ \phi'\in[0,2\pi[\ .$$
This is illustrated in Fig. 3 for the case $(m,n)=(3,2)$ and $(a,b)=(2,-1)$, so
that $({}^\phi_{\sigma_3})=({}^2_3\ {}^1_2) ({}^{\phi'}_{\sigma_3'})$.
Meridian $\gamma'$ and longitude $L'$ of $T'$ are represented on $T$ by $\gamma$ and $L$ as shown.

\begin{figure}[htb]
\vspace{9pt}
\vskip7cm
\epsffile[-140 1 0 0]{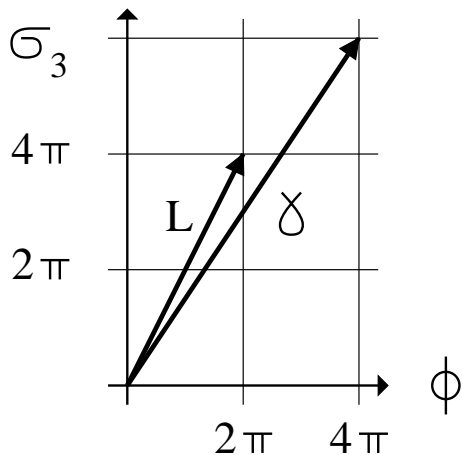}
\caption{\small Surgery with $m=3,n=2$.}
\end{figure}

So a three-dimensional manifold on which a metric
corresponding to fractional magnetic charge (\ref{hanna}) can live
can be constructed by identifying the surfaces $T$ and $T'$ of the two solid tori $S$ and $S'$
by an SL(2,Z) transformation that maps a 
meridian ${\gamma}'$ of $T'$ onto the spiral ${\gamma}$ on $T$.
This is a more general type of Dehn surgery. 
The topologies constructed in this way are the so-called lens spaces $L(m,n)$ \cite{rolfsen}.\footnote{They can be regarded as total spaces of U(1) fibre bundles over $Z_n$ orbifolds of $S^2$ (see e.g. \cite{scott}).}
A lens space $L(m,n+m)$ can be seen to be equivalent 
to $L(m,n)$, so the relatively prime integers $m$ and $n$ can be chosen such that $n<m$.
If $m$ and $n$ are not relatively prime, then a manifold corresponding to magnetic flux
(\ref{hanna}) is the lens space
$L({m\over k},{n\over k})$ where $k$ is the largest common denominator of $m$ and $n$.

So far we have performed surgery on a single $S^1$ fibre of $\Sigma\times S^1$ with $\Sigma=S^2$.
More generally, one can start with an oriented or unoriented
Riemann surface $\Sigma$ of any genus and perform
independent surgery on an arbitrary number of fibers.
In this way, all three-dimensional manifolds that admit a foliation by circles
can be constructed. Those are called ``Seifert manifolds'' \cite{seifert}. 
Within the present ansatz, where the metric $h_{ij}$ depends only on the coordinates on $\Sigma$
and not on $\sigma_3$, these are the only topologies that can occur.

In the case of
the M-brane, one is interested only in orientable Seifert manifolds \cite{bst}.
These can be described as follows \cite{seifert}:
First, the orientation of the fibres is fixed such that if the 
orientation of $\Sigma$ is preserved along a
homology cycle $C$ on $\Sigma$, then the orientation of the fibre is also 
preserved along $C$. If the orientation of $\Sigma$ is reversed along
$C$, then the orientation of the fibre is also reversed along $C$.
Given $\Sigma$, any Seifert manifold is now characterized 
by a number $N$ of ``singular'' fibres,
i.e. fibres around which Dehn surgery has been performed
with integer coefficients $(m_i,n_i),\ i\in\{1,...,N\}$.
One usually normalizes the $m_i$ such that $m_i<n_i$. 
If this normalization is used, then one also needs an $(N+1)$-th fibre  
with surgery coefficients $(m,n)=(b,1)$, where $b$ can be any integer.

{}\section*{4. The phase transition}\scs{4}

Let us now return to the gauged XY model (\ref{kate}) with action
\ba
 {1\over4\pi}\int d^2\sigma\ \{\ {1\over T}(\partial_\alpha\Theta-\tilde A_\alpha)^2\ 
+\ {1\over 4e^2}\tilde F^2\ \}\ .\la{katie}\ea
We recall the relation $R=\kappa^{2\over3}$, the notation $h_{33}=(nR)^2, h_{3\alpha}=(nR)^2A_\alpha$
 and the definitions  
\ba
 T\ =\ {1\over nR^3}\ ,\ \ e^2\ =\ {1\over \gamma nR^3}\ , \ \ \tilde A_\alpha\ =\ nA_\alpha\ ,
\ \ \Theta\ =\ {x_{11}\over R}\ .\la{kimberly}
\ea
It may be helpful in the following to think of this model as
the $\lambda\rightarrow\infty$ limit of an abelian Higgs model with Higgs field 
$\phi\equiv|\phi|e^{i\Theta}$ and Lagrangean
$$
 {1\over4e^2}\tilde F^2\ +\ 
|(\partial_\alpha-i\tilde A_\alpha)\phi|^2\ +\ \lambda(|\phi|^2-{1\over T})^2
\ .$$
In the limit $\lambda\rightarrow\infty$, $|\phi|$ is frozen to ${1\over\sqrt T}$.
For finite $\lambda$, the three-dimensional version of this model would describe a superconductor,
with $\phi$ representing a bound state of two electrons.\footnote{More precisely,
the three-dimensional Abelian Higgs model describes a superconductor of type II
(which has Abrikosov flux tubes) if the ratio of the gauge field mass to the Higgs mass
is small. In the limit $\lambda\rightarrow0$ that we consider here, this ratio becomes zero.}

In the Higgs mechanism,
the Goldstone boson $\Theta$ gets eaten by
the gauge field $\tilde A$ which acquires  mass
$$m_{\tilde A}^2\ =\ {1\over T}\ .$$
As for the superconductor, in system (\ref{katie}) there is then a
Meissner effect: magnetic flux is expelled
from the three-manifold (the membrane), the penetration depth being 
${1\over em_{\tilde A}}\sim{\sqrt T}/e$. 
However, magnetic flux can be trapped in Abrikosov-like flux tubes \cite{abrikosov}.
Those consist of vortex lines of $\Theta$ parallel to the $S^1$, surrounded by magnetic flux. I.e.,
\ba \Theta &=& m\phi\ +\ n\sigma_3\la{rebecca} \\
 \tilde A_\phi &\rightarrow& {m\over r}\ \ \ \hbox{for}\ \ \ r\rightarrow\infty\la{rosi}\\
 \tilde A_\phi &\rightarrow& 0\ \ \ \ \ \hbox{for}\ \ \ r\rightarrow0\ .\la{ruth}
\ea
Here, $(r,\phi,\sigma_3)$ are cylindrical coordinates on $\Sigma\times S^1$
and $\tilde A_\phi$ is the tangential component of the gauge field.
These flux tubes are ``gravitational'' flux tubes, with Ricci curvature proportional to 
the square of the field strength of $\tilde A$
from (\ref{erika}). 

Before discussing them, a remark is in order.
One might think that
vortex lines of $x_{11}$ tear holes into the membrane, in the sense that
the curve defined by $\sigma_3=0,r=\epsilon,\phi\in[0,2\pi[$
does not shrink to a point in embedding space for $\epsilon\rightarrow0$. 
If that was the case, one might perhaps worry that the vortex configurations 
are only allowed in a lattice model as opposed to a continuum field theory.
However, there is a curve that does shrink to a point as $\epsilon\rightarrow0$: 
this is the closed spiral defined by $m\phi+n\sigma_3=0$, 
instead of $\sigma_3=0$. So the embedding of the wrapped membrane in target space
does not have holes.

To confirm the existence and discuss the properties of the magnetic flux tubes, let us first define
\ba B\ \equiv\ {m}-r\tilde A_\phi(r)\ . \la{monika}\ea
Then the field strength is
$$\tilde F\ =\ (\partial_r+{1\over r})\tilde A_\phi\ =\  -{\partial_rB\over r}\ .$$ 
 Action (\ref{katie}) with ansatz (\ref{rebecca}) becomes
$${1\over2}\int {dr\over r}\ \{{(\partial_rB)^2\over 4e^2}\ +\ {B^2\over T}\}\ .$$
The equation of motion of (\ref{katie}) becomes
$$B''(x)\ -\ {1\over x}B'(x)\ -\ B(x)\ =\ 0\ \ \ \ \hbox{with}\ \ x={2e\over\sqrt T}r\ .$$
It is easy to see that it indeed
has smooth solutions that behave like
\ba B(x) &\propto& {\sqrt x}\ e^{-x}\ \hskip20mm \ \hbox{for}\ \ x\rightarrow\infty\\
 B(x) &\sim& {m}(1+{1\over2}x^2\log x)\ \ \ \ \hbox{for}\ \  x\rightarrow0\ ,\ea
so that , from (\ref{monika}), $\tilde A_\phi$ behaves as desired in (\ref{rosi},\ref{ruth}).
From the relation between $x$ and $r$ one sees again that
the radius of the tube outside of which magnetic flux
decays exponentially is of order ${\sqrt T}/e={\sqrt\gamma}$, the penetration depth noted above.

The original gauge field $A_\phi={1\over n}\tilde A_\phi$ behaves like 
$A_\phi\rightarrow{m\over nr}$ at infinity and 
 goes to zero at the vortex center.
Thus the enclosed magnetic flux is 
$$\int F\ =\ \oint A\ =\ 2\pi{m\over n}\ .$$

But as we have seen, this fractional magnetic charge
 changes the topology of the manifold; the new topology can be obtained by performing
Dehn surgery with coefficients $(m,n)$ along the vortex line as discussed in the 
previous section.
Thus we are led to identify the magnetic flux tubes with the
tubes labelled by $m$ and $n$ that are cut out of $\Sigma\times S^1$
and glued back in in the process of constructing $M$ by surgery!
We conclude that the Dehn surgery tubes ``are really there'', in the sense that
if the topology of the wrapped M-brane $M$ is different from $\Sigma\times S^1$,
then the topological defects and the corresponding Ricci curvature are
confined to the inside of tubes of size $\sim{\sqrt\gamma}$.

What is the action of such a ``surgery tube''?
First, the gauge field contribution to the action is finite (and independent of the undetermined
parameter $\gamma$), namely of order
$${1\over e^2}\int d^2\sigma\ \tilde F^2\ \sim\ ({1\over e^2})(A_{tube})({2\pi m\over A_{tube}})^2\
\sim\ {m^2\over e^2A_{tube}}\ \sim\ {m^2\over T}\ .$$
Here we have used the fact that magnetic flux $\int\tilde F =2\pi m$ is, roughly, distributed over
an area of size $A_{tube}\sim T/e^2$.

On the other hand, the contribution of $\Theta$ to the action of a flux tube
 diverges;\footnote{This differs
from the situation in the abelian Higgs model with {\it finite} $\lambda$, where
 both the gauge field and the Higgs contributions to the action
are well-known to be finite.}
the divergence arises from
the region $r\rightarrow0$ (where the gauge field vanishes):
$$S_\Theta\ =\ {1\over 2T}\int_a {dr\over r}(\partial_\phi\Theta)^2\ =\ {m^2\over2T}
\vert\log a\vert\ .$$
Here the divergent integral had to be cut off by a
short-distance cutoff $a$.

This logarithmic divergence of the action means that the vortex density has an anomalous dimension ${1\over2}m^2\kappa^2$.
It is responsible for the well-known
Kosterlitz-Thouless phase transition \cite{kt} which occurs when this dimension is two:
intuitively, the density of possible locations of a vortex (e.g.
of lattice sites in a lattice model) is proportional 
to $1\over a^2$,
so the expectation value of the vortex density is proportional to
$${1\over a^2}\ \exp\{-{m^2\over2T} |\log a|\}\ \exp\{-{m^2\over T}\}\ \sim\ a^{-2+{m^2\over2T}}
\ \exp\{-{m^2\over T}\}\ .$$
As $a\rightarrow0$, this is zero for $T<{m^2\over4}$ and infinite for $T>{m^2\over4}$.
Thus we conclude that Dehn surgery tubes condense at critical temperature
$$T_c\ =\ {m^2\over4}\ .$$
This is the same result as one gets with the standard analysis of the
XY model without gauge field, 
which maps the Coulomb gas of vortices onto a sine-Gordon model
at inverse temperature (see, e.g., \cite{zinn}).
Note however that the transition here takes place not because of a
long-range Coulomb attraction between the vortices (which is removed by the gauge field),
but because of the short-distance divergence of the action.
Since $A\rightarrow0$ at the vortex center, the critical temperature and the order of the phase 
transition (which is second order, but with the well-known two-dimensional peculiarities)
are independent of the presence of the gauge field.\footnote{This 
again differs from the case of finite $\lambda$. 
There the second-order phase transition of the ungauged Higgs model is 
known to be changed to a first-order transition by the coupling to the gauge field.}

In the case $n=1$ one has $T={1\over R^3}\sim{1\over\kappa^2}$. So 
the critical temperature corresponds to a critical string coupling constant
\ba
\kappa_c\ \sim\ {2\over m}\ ,\la{susi}\ea
whose value is independent of the undetermined parameter $\gamma$.
For $n\neq1$, one would get $\kappa_c\sim{2\over m{\sqrt n}}$.
We will comment below on the fact that $m$ and $n$ do not enter on equal par in this formula.

Up to now, model (\ref{gabi}) has been discussed in flat two-dimensional space.
It remains to discuss how the phase transition changes
when the model is coupled to dynamical gravity. Now, as pointed out above,
the mechanism responsible for the phase 
transition is independent of the presence of the gauge field.
So it suffices to study 
the effect of gravity on the Kosterlitz-Thouless transition in an 
ordinary, i.e. ungauged XY model. But this has been investigated previously \cite{gross},
at least in the case where the compact scalar is the only target-space coordinate. It is known
that the transition is not changed qualitatively, although
the one-loop beta functions get multiplied by an overall factor \cite{schmisine,kkp}.
For this reason we assume that the critical values $\kappa_c$ are not 
changed by coupling model (\ref{katie}) to gravity either.

There is a point that should perhaps be adressed here.
What is the meaning of phase transitions,
 beta functions and scale-dependent coupling constants in a theory
with gravity, where the scale itself is integrated over?
The answer is known from ``old'' (i.e., perturbative) string theory:
the scale factor $\phi$ of the two-dimensional world-sheet metric on $\Sigma$
becomes the time coordinate of target space.
Thus, overall scale transformations, i.e., renormalization group transformations on the world-sheet
 correspond to time translations in space-time.\footnote{This 
conclusion is,
to the author's knowledge, first explicitly discussed in \cite{das}.}
Renormalization group trajectories then correspond to classical solutions of string theory
(see \cite{st} for a discussion along these lines).

{}\section*{5. Open ends}\scs{5}

\vskip5mm
$\underline{\hbox{Conclusion and speculation}}$\vskip2mm

Let us summarize the conclusions.
The three-dimensional topologies of M-branes that are wrapped around the circular eleventh dimension
can be constructed from the manifolds
$\Sigma\times S^1$, where $\Sigma$ is a Riemann surface, by performing Dehn surgery with
independent coefficients $(m_i,n_i)$ along an arbitrary number of tubes that run around the $S^1$.

The world-brane theory of the wrapped M-brane has been argued to
include Maxwell theory coupled to the eleventh dimension of M-theory.
The Dehn surgery tubes correspond to magnetic flux tubes in this model.
For large string coupling constant, they have been argued to be irrelevant,
at least in the model based on the simplifying ansatz (\ref{sima}).
The dominant topologies are therefore $\Sigma\times S^1$, and
the sum over topologies of the membrane reduces to a sum over the genus  of $\Sigma$.
The gauge theory is Higgsed; the eleventh dimension of M-theory
gets ``eaten'' by the $h_{3\alpha}$ component of the auxiliary metric.
What remains is a standard string; the only remnant of the
path integral over $x_{11}$ is the sum over the wrapping number $n$.

One may speculate that, as
in the standard Kosterlitz-Thouless transition, 
the radius of the circle parametrized by $x_{11}$, 
and therefore the string coupling constant $\kappa$ flows to $\infty$ in the infrared
(``infrared'' = ``late times'', according to the discussion at the end of section 4).

At the largest critical value of $\kappa$ in (\ref{susi}) 
there has been argued to occur a phase transition due to the condensation 
of flux tubes. The typical topology of the wrapped membrane 
is then a complicated Seifert manifold.
The gauge theory is in a Coulomb phase
and one may speculate that $\kappa$ flows to zero. 
The existence of such a phase transition in the world-brane theory of the 
wrapped M-brane, as defined in section 2, is our main conclusion. 

Let us now add some comments.
Formula (\ref{susi}) has been derived for $n=1$. As mentioned, for
$n>1$ the winding numbers $m$ and $n$ of $x_{11}$ in $\phi$- and $\sigma_3$-direction
would enter asymmetrically.
This is a consequence of ansatz (\ref{sima}):
since $h_{33}$ is constant, winding of $x_{11}$ around the $\sigma_3$ direction
cannot contribute to the logarithmic divergence of the action and thereby to the critical temperature.
(The factor 
$\sqrt n$ in (\ref{susi}) is only a consequence of the overall factor $nR$ in 
front of the action (\ref{gabi})).

If instead of making ansatz (\ref{sima}) one
introduces $h_{33}(\sigma)$ as an independent field, then
one expects an $SL(2,Z)$ multiplet of flux tubes labelled by $(m,n)$,
due to modular invariance on the torus that is parametrized by $\phi$ and $\sigma_3$. 
In particular, one expects an $SL(2,Z)$ covariant expression for the coefficient of the logarithmic divergence
of the vortex action.

We must leave it to future work to investigate this case.
We must also leave it to the future to discuss the nonabelian generalization of our theory
that describes wrapped $p$-branes with $p>2$.

\vskip5mm\noindent $\underline{\hbox{Confined momentum?}}$\vskip2mm

In ansatz (\ref{clara}) we have excluded ``by hand'' oscillations in $\sigma_3$-direction 
from the allowed configurations of $\vec x(\sigma)$. Could this restriction arise automatically?

In \cite{wittencom} it has been suggested (in the context of a proposal 
to explain the vanishing of
the cosmological constant) that the Kaluza-Klein gauge field that arises upon dimensional
reduction of four-dimensional gravity to three dimensions along an $S^1$ might be linearly confining;
the confined charge would be Kaluza-Klein momentum in the $S^1$-direction.
Perhaps something similar can be suggested in order to justify ansatz (\ref{clara}).
Perturbing one of the coordinates $\vec x$ in ansatz (\ref{clara}) by
an oscillation proportional to $e^{ik\sigma_3}$
amounts to introducing a new two-dimensional field (a Kaluza-Klein mode) in action (\ref{gabi}),
whose charge and mass are equal to $k$.
Confinement of electric charge in the two-dimensional gauge theory in (\ref{gabi}) would indeed mean
that all the oscillators $e^{ik\sigma_3}$ can be ignored at large scales.
So the ``wrapping of branes'' might have an interpretation in terms of
confinement of world-brane momentum.

We plan to discuss these and other issues left open here in future work.

{}

\end{document}